\documentclass{aa}
\usepackage{graphics}

\begin{document}
\title{Thermal infrared observations of the Hayabusa spacecraft target asteroid
       \object{25143~Itokawa}
       \thanks{Based on observations collected at the European
               Southern Observatory, Chile;
               ESO, No.\ 73.C-0772}}
\author{T.\ G.\ M\"uller\inst{1}
	\and
	T.\ Sekiguchi\inst{2}
        \and
	M.\ Kaasalainen\inst{3}
        \and
        M.\ Abe\inst{4}
        \and
        S.\ Hasegawa\inst{4}
	}

\authorrunning{M\"uller et al.}
\titlerunning{Itokawa in the thermal infrared}

\offprints{T.\ G.\ M\"uller}

\institute{%
    Max-Planck-Institut f\"ur extraterrestrische Physik,
    Giessenbachstra{\ss}e, 85748 Garching, Germany;
    \email{tmueller@mpe.mpg.de}
 \and
    National Astronomical Observatory of Japan, 2-21-1 Osawa, Mitaka, Tokyo 181-8588, Japan;
    \email{t.sekiguchi@nao.ac.jp}
 \and
    Department of mathematics and statistics,
    Gustaf Hallstromin katu 2b, P.O. Box 68,
    FIN-00014 University of Helsinki, Finland;
     \email{mjk@rni.helsinki.fi}
 \and
    Institute of Space and Astronautical Science,
    Japan Aerospace Exploration Agency,
    3-1-1 Yoshinodai, Sagamihara, Kanagawa 229-8510, Japan;
     \email{abe@planeta.sci.isas.jaxa.jp};
     \email{hasehase@isas.jaxa.jp}
    }

\date{Received / Accepted for publication 29/Aug/2005}

\abstract{We obtained N- and Q-band observations of the Apollo-type asteroid
          \object{25143~Itokawa} during its close Earth approach in July 2004 with
          TIMMI2 at the ESO 3.6\,m telescope. Our photometric measurement, in
          combination with already published data, allowed
          us to derive a radiometric effective diameter of 0.32$\pm$0.03\,km
          and an albedo of 0.19$^{+0.11}_{-0.03}$ through a thermophysical model.
          This effective diameter corresponds to a slightly asymmetrical
          and flattened ellipsoid of the approximate size of
          520($\pm$50)$\times$270($\pm$30)$\times$230($\pm$20)\,m, based
          on the Kaasalainen et al. (\cite{kaasalainen05}) shape model.
	  Our studies show that the thermal observations lead to size
          estimates which are about 15\% smaller than the radar
          results (Ostro et al.\ \cite{ostro05}), slightly outside
          the stated radar uncertainties of $\pm$10\%.
          We determined a rather high thermal inertia of
          750\,J\,m$^{-2}$\,s$^{-0.5}$\,K$^{-1}$. This is an indication
          for a bare rock dominated surface, a thick dust regolith can be
          excluded as well as a metallic surface.
	  From our data we constructed a 10.0\,$\mu$m thermal
          lightcurve which is nicely matched in amplitude and phase
	  by the shape and spin vector solution in combination with
          our TPM description.
	  The assumed S-type bulk density in combination with radiometric
	  size lead to a total mass estimate of
	  $4.5^{+2.0}_{-1.8} \cdot 10^{10}$\,kg.
  \keywords{Minor planets, asteroids -- Radiation mechanisms: Thermal --
            Infrared: Solar system}
}

\maketitle

\section{Introduction}

  The Near-Earth asteroid (NEA) 25143~Itokawa (1998\,SF36) is
  the target for the Japanese Hayabusa (MUSES-C) sample return
  mission. The spacecraft will arrive at the asteroid in summer
  2005 and hover close to the surface for about 3 months before
  it will collect surface samples which will be brought back
  to Earth in June 2007.
  
  Several ground-based observing campaigns
  took place during the last years to derive various properties
  of Itokawa. E.g., Binzel et al. (\cite{binzel01}) concluded from
  visible and near-infrared spectroscopic measurements that
  the spectral characteristics (S(IV)-type) match the LL ordinary
  chondrite class meteorites. Ostro et al.\ (\cite{ostro04}; \cite{ostro05})
  report on delay-Doppler images obtained at Arecibo and Goldstone,
  resulting in size, shape, radar albedo and surface roughness
  estimates. Kaasalainen et al.\ (\cite{kaasalainen03}; \cite{kaasalainen05})
  determined through lightcurve inversion techniques a high quality 
  shape and pole solution together with a period of
  $P_{sid}= 12.13237\pm 0.00008$\,h.
  Sekiguchi et al.\ (\cite{sekiguchi03}) observed \object{Itokawa}
  at thermal infrared wavelengths and used the powerful radiometric
  technique (e.g.\ Harris \& Lagerros \cite{harris02}) to determine
  a size of 0.35$\pm$0.03\,km and an albedo of 0.23($+$0.07, $-$0.05).
  M\"uller et al.\ (\cite{mueller04}) reported on thermal property
  studies of \object{Itokawa} based on multi-epoch thermal infrared
  photometric data. But their data set did not allow to find
  a unique solution for the size, thermal inertia and roughness.
  Based on the radar size, they derived a thermal inertia between
  5 and 10 times that of the Moon.
  
  Here, we used all available thermal infrared data together with own 
  observations, taken during a close approach in July 2004 (Section
  \ref{sec:obs}).
  The  modelling and the derivation of thermophysical properties
  of \object{Itokawa} was then performed through
  the well-established, tested and frequently used thermophysical
  model (TPM) by Lagerros
  (\cite{lagerros96}; \cite{lagerros97}; \cite{lagerros98}; see
  Sects.~\ref{sec:tpm}, \ref{sec:mod} and \ref{sec:modnew}). The results are
  discussed in the context of the already
  known properties of \object{Itokawa} (Sect.~\ref{sec:discussion})
  and conclusions are drawn in Sect.~\ref{sec:conclusion}.
  
  
\section{Observations and Data Reduction}
\label{sec:obs} 

  We combined the observations from Sekiguchi et al.\ (\cite{sekiguchi03})
  (data set \#1 in Table~\ref{tbl:obslog}), with the data set from Delbo (\cite{delbo04}) (data set \#2),
  and our own observations (data set \#3).
  All measurements were taken with the
  TIMMI2 instrument (K\"aufl et al. \cite{kaeufl03}) at the ESO La Silla
  3.6\,m telescope. Table~\ref{tbl:obslog} summarises the observing geometries for
  all 20 measurements.

  \begin{table*}[h!tb]
  \begin{center}
    \caption{Summary of TIMMI2 observations of asteroid \object{25143~Itokawa}.
             The phase angles are negative before opposition and positive after.
	     We added data from Sekiguchi et al.\ (\cite{sekiguchi03})
	     and Delbo (\cite{delbo04}).}
             \label{tbl:obslog}
    \begin{tabular}{rlllllll}
      \hline
      \hline
      \noalign{\smallskip}
           & \multicolumn{2}{c}{Mid-Time} & Filter & r    & $\Delta$ & $\alpha$     & \\
      No   & \multicolumn{2}{c}{(Day UT)} & Band   & [AU] & [AU]     & [$^{\circ}$] & Remarks \\
      \noalign{\smallskip}
      \hline
      \noalign{\smallskip}
      01 & 2001/Mar/14 & 05:50 & N11.9 & 1.059232 & 0.073897 &  +27.54 & Sekiguchi et al. (\cite{sekiguchi03}) \\
      \noalign{\smallskip}
      02 & 2001/Apr/08 & 09:27 & N11.9 & 0.983221 & 0.053606 & 108.33 & Delbo (\cite{delbo04}) \\
      03 & 2001/Apr/08 & 09:42 & N10.4 & 0.983198 & 0.053633 & 108.35 & and priv.\ comm. \\
      04 & 2001/Apr/08 & 10:01 & N12.9 & 0.983169 & 0.053667 & 108.37 & " \\
      05 & 2001/Apr/08 & 10:18 & N8.9  & 0.983142 & 0.053698 & 108.38 & " \\
      06 & 2001/Apr/08 & 10:34 & N11.9 & 0.983117 & 0.053728 & 108.40 & " \\
      07 & 2001/Apr/09 & 09:28 & N12.9 & 0.981024 & 0.056409 & 109.93 & " \\
      08 & 2001/Apr/09 & 09:45 & N9.8  & 0.980999 & 0.056441 & 109.95 & " \\
      09 & 2001/Apr/09 & 10:03 & N10.4 & 0.980972 & 0.056475 & 109.96 & " \\
      10 & 2001/Apr/09 & 10:18 & N11.9 & 0.980949 & 0.056504 & 109.98 & " \\
      11 & 2001/Apr/09 & 10:32 & N11.9 & 0.980928 & 0.056530 & 109.99 & " \\
      \noalign{\smallskip}
      12 & 2004/Jul/01 & 06:03 & N1    & 1.028243 & 0.020164 &  -54.63 & this work \\ 
      13 & 2004/Jul/01 & 06:19 & N1    & 1.028279 & 0.020193 &  -54.56 & " \\         
      14 & 2004/Jul/01 & 06:36 & N1    & 1.028318 & 0.020224 &  -54.49 & " \\         
      15 & 2004/Jul/01 & 06:54 & N1    & 1.028359 & 0.020257 &  -54.41 & " \\         
      16 & 2004/Jul/01 & 07:16 & N2    & 1.028409 & 0.020298 &  -54.31 & " \\         
      17 & 2004/Jul/01 & 07:36 & N2    & 1.028454 & 0.020335 &  -54.22 & " \\         
      18 & 2004/Jul/01 & 07:53 & N12.9 & 1.028492 & 0.020367 &  -54.15 & " \\         
      19 & 2004/Jul/01 & 08:09 & N12.9 & 1.028529 & 0.020397 &  -54.08 & " \\         
      20 & 2004/Jul/01 & 08:37 & Q1    & 1.028592 & 0.020450 &  -53.95 & " \\         
      \noalign{\smallskip}
    \noalign{\smallskip}
    \hline
    \end{tabular}
  \end{center}
  \end{table*}
  
  A standard chopping and nodding technique was utilized for all
  observations to reduce the atmospheric and telescope background
  emission. Chop and nod throws were 10$^{\prime \prime}$,
  respectively. For the imaging observations, a pixel scale of
  0.2$^{\prime \prime}$ was chosen, and on source integration
  times were 22\,min (obs \#1), 10-13\,min (obs \#2 to \#11),
  7.5\,min (obs \#12 to \#19), and 25\,min (obs \#20).
  
  Both already published data sets, from Sekiguchi et al. (\cite{sekiguchi03}) and
  from Delbo (\cite{delbo04}), were re-calibrated using the official
  central filter wavelengths in combination with the flux densities of the
  corresponding stellar models (see Tbl.~\ref{tbl:starcal}).

\begin{figure*}[h!tb]
    \begin{center}
      \resizebox{6cm}{!}{\includegraphics{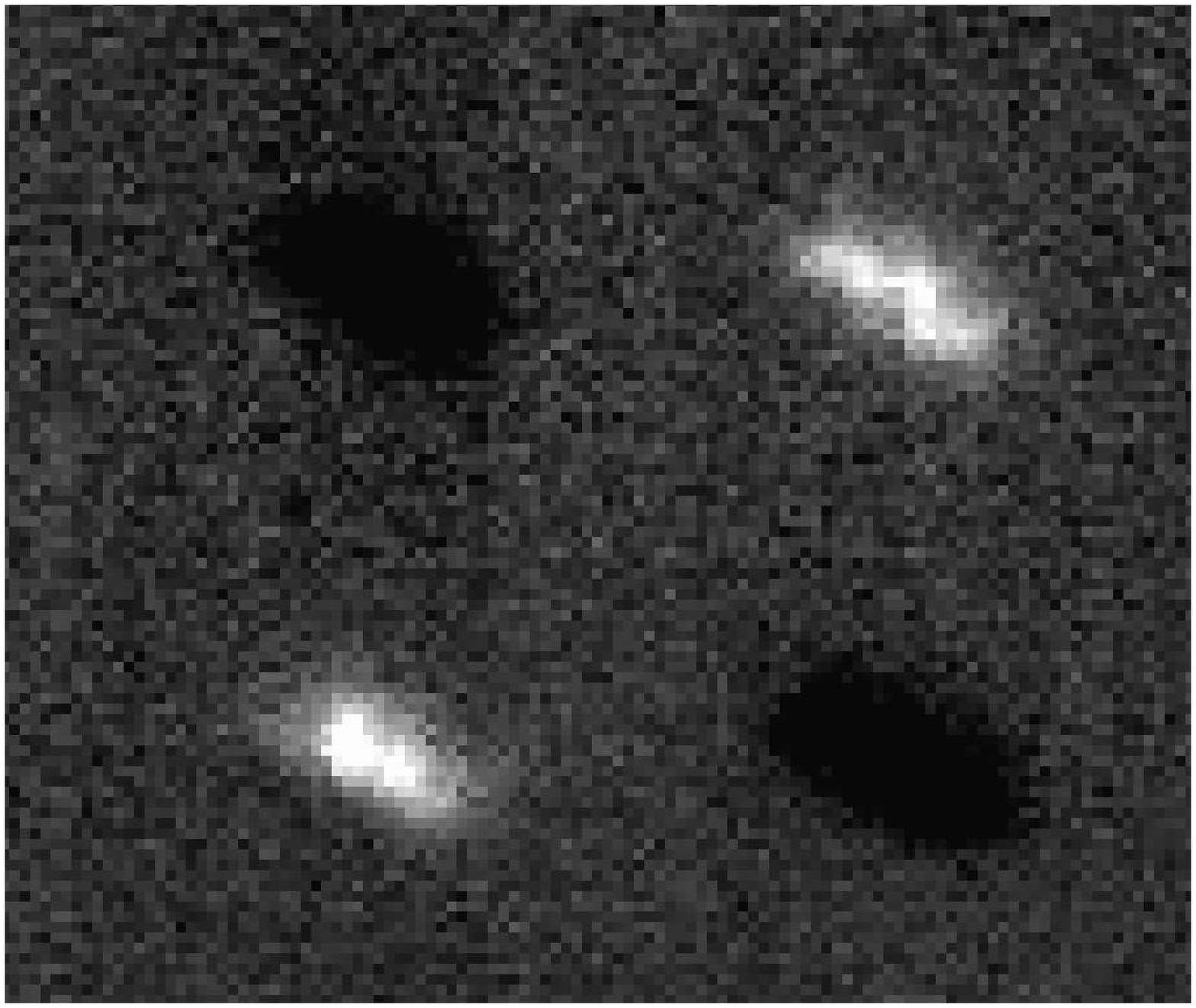}}
      {\large $\rightarrow$}
      \resizebox{6cm}{!}{\includegraphics{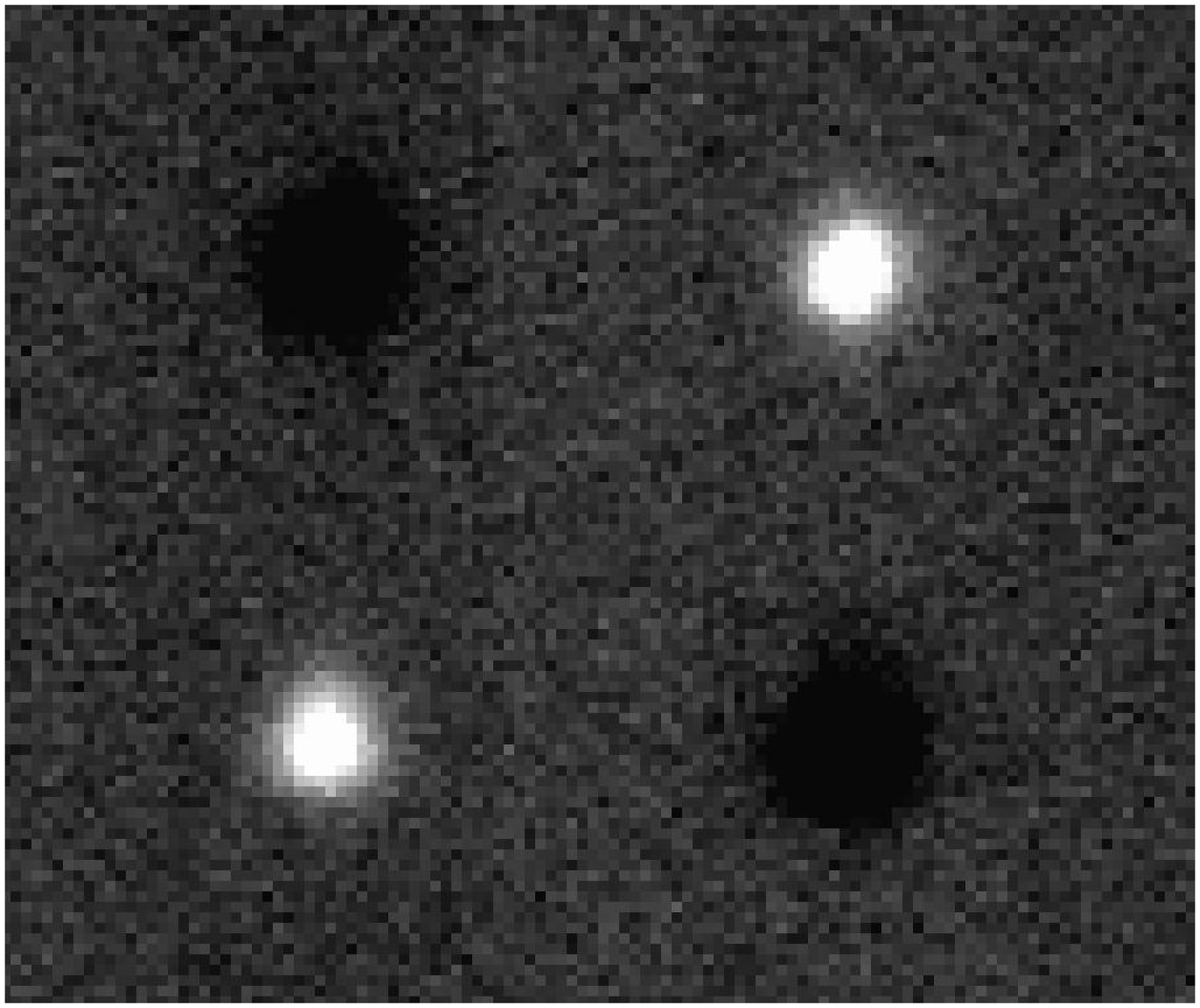}}
      \resizebox{6cm}{!}{\includegraphics{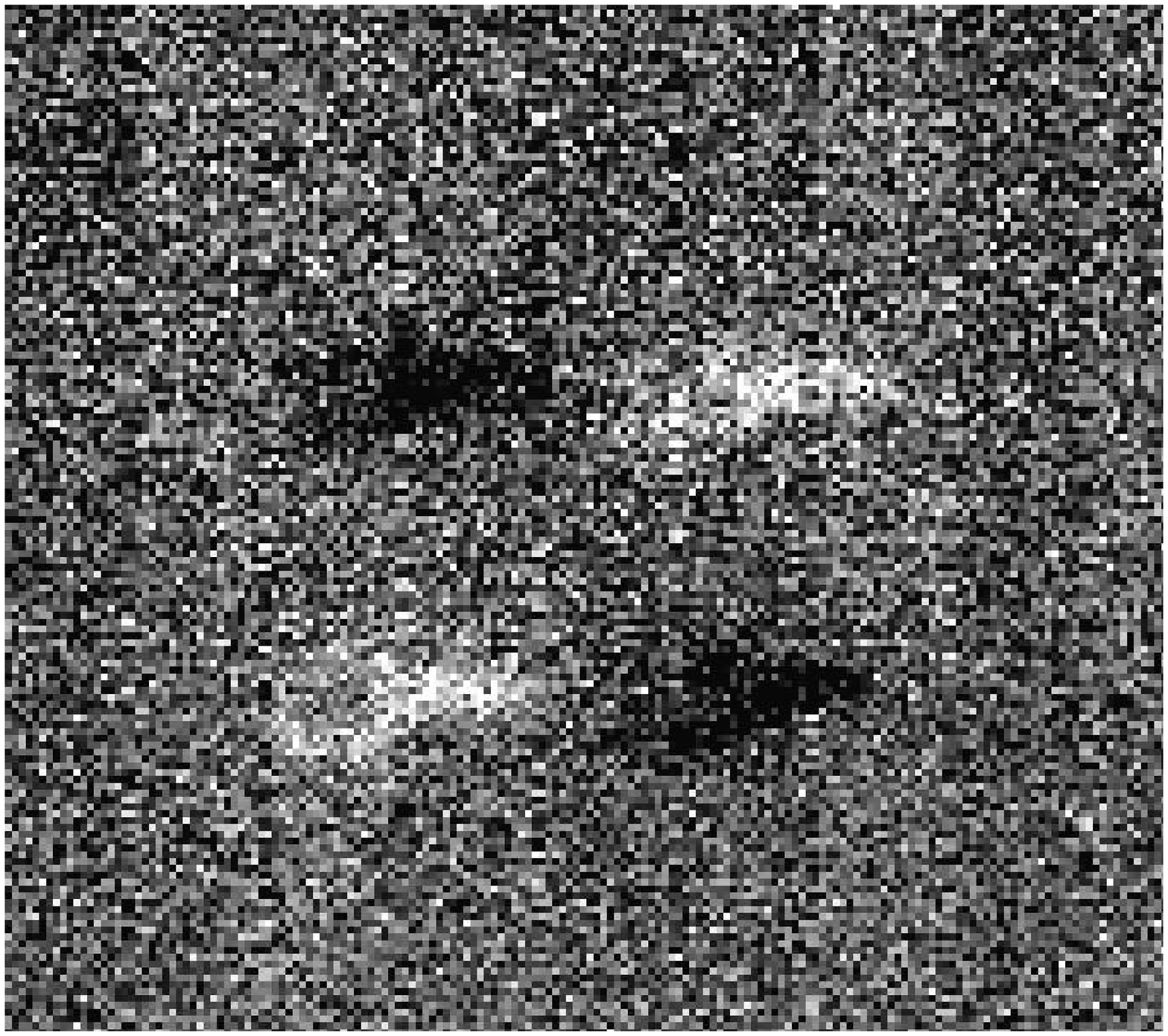}}
      {\large $\rightarrow$}
      \resizebox{6cm}{!}{\includegraphics{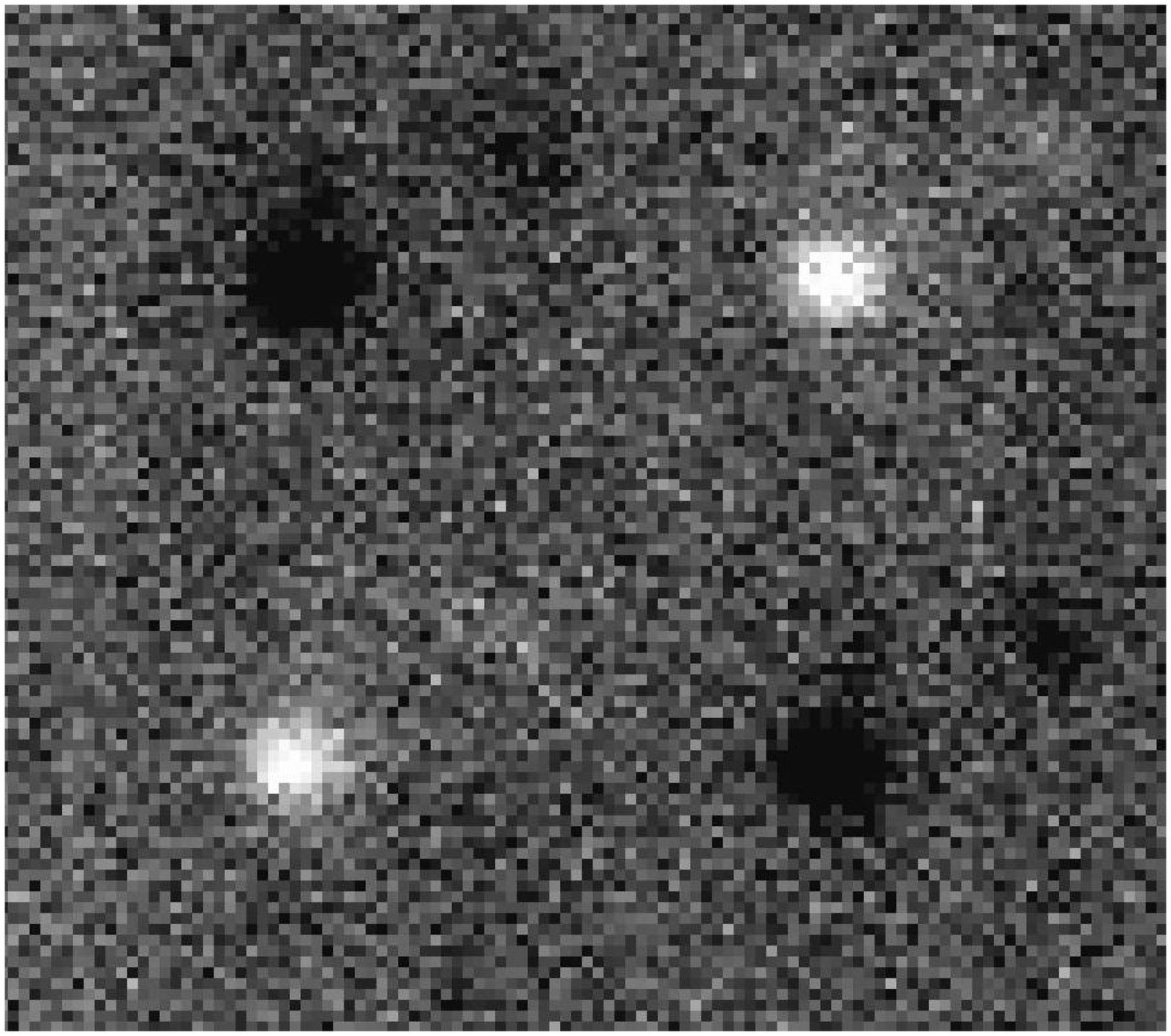}}
     \caption{Pipeline-processing in comparison with re-reduced asteroid images.
                  Top: Itokawa measurement \#12 (Tbl.~\ref{tbl:obslog}), N1-filter,
	               7.5\,min integration time, left: pipeline-product,
		       right: the 60 raw images were shifted by 
		       1.0 pixels per minute in x-direction and by -1.0
		       pixels per minute in y-direction and then co-added.
               Bottom: Itokawa measurement \#20 (Tbl.~\ref{tbl:obslog}), Q1-filter,
	               25.8\,min integration time, left: pipeline-product,
		       right: the 234 raw images were shifted by
		       0.8 pixels per minute in x-direction and by 0.2
		       pixels per minute in y-direction and then co-added.
       \label{fig:elongated}}
    \end{center}
  \end{figure*}

  A more detailed data reduction and calibration was performed on
  our own data set. Unfortunately, the telescope tracking on
  25143~Itokawa was not perfect in the 2004 observing run. This
  had the consequence that the pipeline-reduced images showed elongated
  sources (see left side of Fig.~\ref{fig:elongated}).
  Therefore, we took the raw images and processed them in
  a pipeline-like manner together with a fixed x-y-shift rate of
  typically 0.5 to 1.0 pixels/minute. A centroid determination in
  combination with a standard shift-and-add technique was not possible
  due to the faint sources which are not always visible on individual
  raw frames. The results of this process are illustrated on the right
  side of Fig.~\ref{fig:elongated}. 
  
  \begin{table*}[h!tb]
  \begin{center}
    \caption{Monochromatic flux densities in [Jy] of the stellar calibrators
             at the TIMMI2 central filter wavelengths.
  Note: The key wavelengths were taken from
  {\tt http://www.ls.eso.org/lasilla/sciops/3p6/timmi/Filters}.
  The model fluxes were taken from
  {\tt http://www.iso.vilspa.esa.es\-/users\-/expl\_lib\-/ISO\-/wwwcal\-/isoprep\-/cohen/\-templates/}
  and interpolated to the central filter wavelengths.}
             \label{tbl:starcal}
  \begin{tabular}{lllllllll}
           & \multicolumn{8}{c}{Filter/Wavelength [$\mu$m]} \\
      Star & N1    & N8.9  & N9.8  & N10.4 & N2    & N11.9 & N12.9 & Q1 \\
           & 8.70  & 8.73  & 9.68  & 10.38 & 10.68 & 11.66 & 12.35 & 17.72 \\
    \noalign{\smallskip}
    \hline
    \noalign{\smallskip}
  HD47105  &  9.16 &  9.08 &  7.47 &  6.52 &  6.17 &  5.19 &  4.58 &  2.27 \\
  HD123139 & 69.97 & 69.51 & 57.37 & 50.09 & 47.61 & 40.08 & 35.59 & 17.63 \\
  HD196171 & 25.58 & 25.42 & 20.98 & 18.32 & 17.41 & 14.65 & 13.01 &  6.45 \\
    \noalign{\smallskip}
    \hline
  \end{tabular}
  \end{center}
  \end{table*}

  We applied standard aperture photometry on the tracking corrected
  images with same apertures on the stars and asteroids. The aperture
  radii were chosen with respect to the growth curves
  (for details see e.g.\ Delbo \cite{delbo04}). All four on-array signatures
  of the sources were used for the flux calibration.
  Colour differences between stars and \object{25143~Itokawa}
  were negligible (about 1-3\,\%) for the used filters in combination
  with the atmospheric transmission at La Silla
  ({\tt http://www.ls.eso.org/\-lasilla/\-sciops/\-3p6/\-timmi/\-html/\-Atmospheric\-Transm.html}).
  
  The observational results are summarised in Tbl.~\ref{tbl:obsres}.

  \begin{table}[h!tb]
    \begin{center}
    \caption{Summary of TIMMI2 observational results for asteroid \object{25143~Itokawa}.
             Note: The Sekiguchi et al.\ (\cite{sekiguchi03}) and Delbo (\cite{delbo04})
	     fluxes and errors have been recalculated for the true central
	     filter wavelengths, using the corresponding stellar fluxes.
	     Measurements from April 9, 2001 (marked with $\star$) were taken
	     under less favorable conditions in comparison with data from 
	     April 8 (M.\ Delbo, priv.\ comm.). Our new observations are listed
	     under \#12 until \#20.
             \label{tbl:obsres}} 
    \begin{tabular}{clrlll}
      \hline
      \hline
      \noalign{\smallskip}
           &        & $\lambda_c$ & FD   & $\sigma_{err}$  &  \\
      No   & Filter & [$\mu$m]    & [Jy] & [Jy]            & Remarks \\
      \noalign{\smallskip}
      \hline
      \noalign{\smallskip}
      01 & N11.9 & 11.66 & 0.264 & $\pm$0.044 & Sekiguchi et al.\ (\cite{sekiguchi03})\\
         &       &       &       &            & (re-calibrated)                    \\
      \noalign{\smallskip}
      02 & N11.9 & 11.66 & 0.164 & $\pm$0.021 & Delbo (\cite{delbo04}) \\
      03 & N10.4 & 10.38 & 0.144 & $\pm$0.018 & and priv.\ comm. \\
      04 & N12.9 & 12.35 & 0.170 & $\pm$0.022 & (re-calibrated) \\
      05 & N8.9  &  8.73 & 0.086 & $\pm$0.022 & " \\
      06 & N11.9 & 11.66 & 0.149 & $\pm$0.019 & " \\
      07 & N12.9 & 12.35 & 0.258 & $\pm$0.032 & " ($\star$) \\
      08 & N9.8  &  9.68 & 0.108 & $\pm$0.016 & " ($\star$) \\
      09 & N10.4 & 10.38 & 0.169 & $\pm$0.027 & " ($\star$) \\
      10 & N11.9 & 11.66 & 0.242 & $\pm$0.030 & " ($\star$) \\
      11 & N11.9 & 11.66 & 0.193 & $\pm$0.028 & " ($\star$) \\
      \noalign{\smallskip}
      12 & N1    &  8.73 & 1.92 & $\pm$0.15 & this work \\  
      13 & N1    &  8.73 & 1.97 & $\pm$0.16 & " \\	   
      14 & N1    &  8.73 & 1.75 & $\pm$0.14 & " \\	   
      15 & N1    &  8.73 & 1.67 & $\pm$0.13 & " \\	   
      16 & N2    & 10.68 & 1.94 & $\pm$0.14 & " \\	   
      17 & N2    & 10.68 & 1.89 & $\pm$0.13 & " \\	   
      18 & N12.9 & 12.35 & 2.17 & $\pm$0.13 & " \\	   
      19 & N12.9 & 12.35 & 1.80 & $\pm$0.11 & " \\	   
      20 & Q1    & 17.72 & 2.49 & $\pm$0.50 & " \\	   
      \noalign{\smallskip}
    \noalign{\smallskip}
    \hline
    \end{tabular}
    \end{center}
  \end{table}

  The error values of observations \#12 to \#20 include the uncertainties
  of the calibration star models (3\,\% in N-band and 4\,\% in Q-band),
  the aperture photometry error (N1: 3\,\%, N2: 2\,\%, N12.9: 2\,\%,
  Q1: 10-20\,\%) and an error for the flat-field residuals (about 3-6\,\%
  in N and about 10-15\,\% for Q-band, depending on the relative placement
  of the calibrator and the asteroid on the TIMMI2 array). The error
  calculations of observations \#1 to \#11 are described in the
  corresponding references.


\section{Thermophysical model description}
\label{sec:tpm}

  We applied the TPM by Lagerros
  (\cite{lagerros96}; \cite{lagerros97}; \cite{lagerros98})
  to all 20 measurements from Tbl.~\ref{tbl:obsres} to investigate
  the physical and thermal properties of \object{Itokawa}.
  On the large scale, the TPM considers the asteroid size, the global
  shape and spin vector and the actual observing and illumination
  geometry at the time of an observation. On the small micrometer
  scale, the TPM takes into account the
  reflected, absorbed and emitted energy, and also the heat
  conduction into the surface regolith. The albedo and emissivity
  control the energy balance and thereby the surface temperature.
  The thermal inertia in combination with the rotation period and
  the orientation of the spin vector influence the diurnal temperature
  variations. As a result, the thermal inertia is strongly connected
  to the interpretation of mid-IR observations, namely when comparing
  before and after opposition observations at large phase angles with
  very different temperatures of the terminator.
  Moreover, the thermal inertia determines the
  amplitude of the thermal lightcurve for a given aspect angle. 
  The TPM beaming model,
  described by $\rho$, the r.m.s.\ of the surface slopes and $f$,
  the fraction of the surface covered by craters, accounts for the
  non-isotropic heat radiation, noticeable at phase angles close
  to opposition. But it also influences
  the shape of the spectral energy distribution in the mid-IR. Detailed
  considerations of the $\Gamma$, $\rho$ and $f$ influences
  for various observing geometries and wavelengths are discussed
  in e.g., M\"uller (\cite{mueller02a}) or Dotto et al.\ (\cite{dotto00}).

  An $H_{\rm{V}}$-value of 19.9 (Kaasalainen et al. \cite{kaasalainen03};
  M. Abe, priv.\ comm.; Sekiguchi et al. \cite{sekiguchi03}) and a
  $G$-value of 0.21 (Abe et al. \cite{abe02a}, \cite{abe02b}) was used
  to describe the visual brightness of \object{Itokawa}.
  We assumed a constant emissivity of 0.9 at all wavelength.

  \begin{figure}[h!tb]
    \begin{center}
      \resizebox{\hsize}{!}{\includegraphics{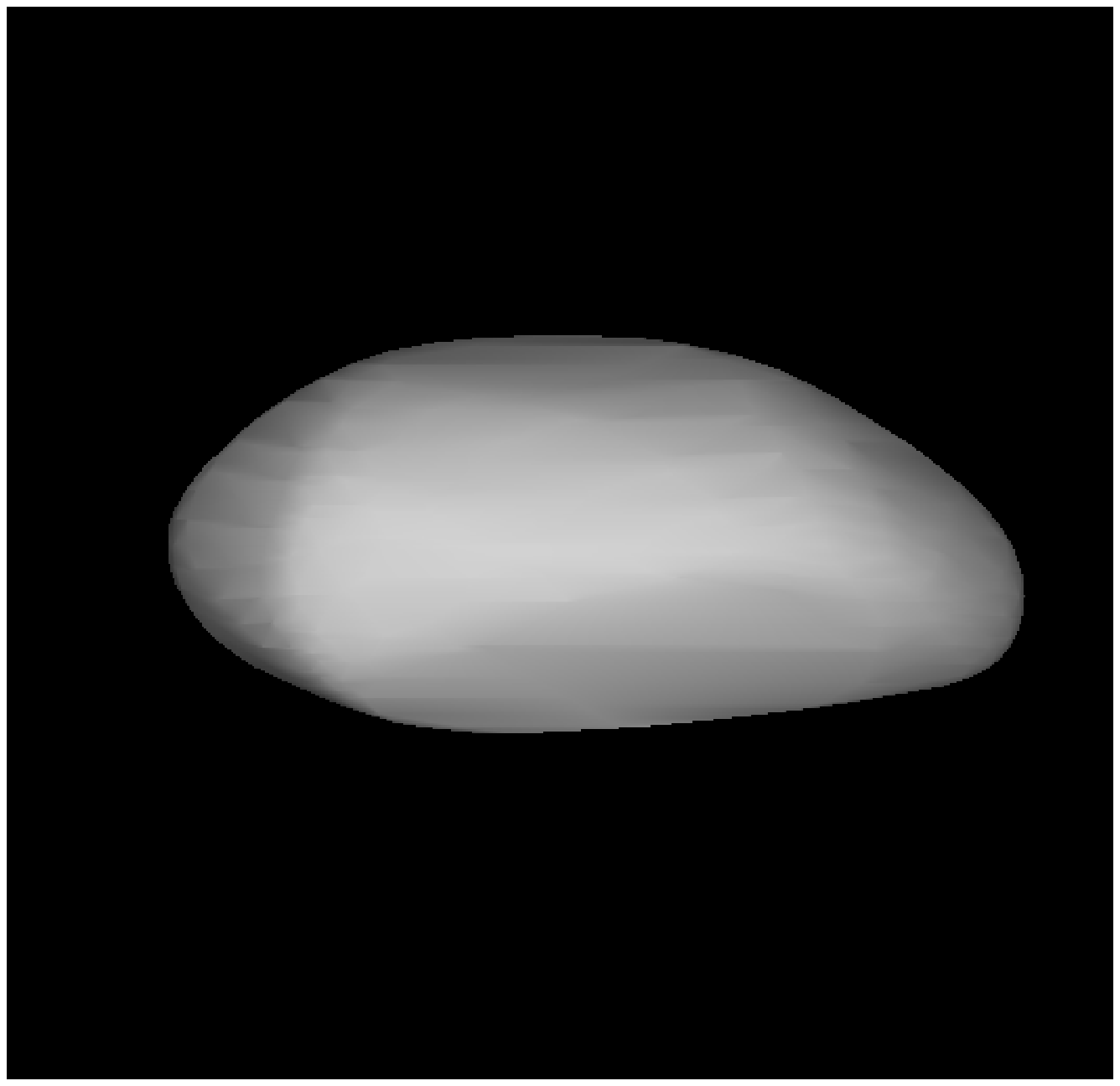}}
      \resizebox{\hsize}{!}{\includegraphics{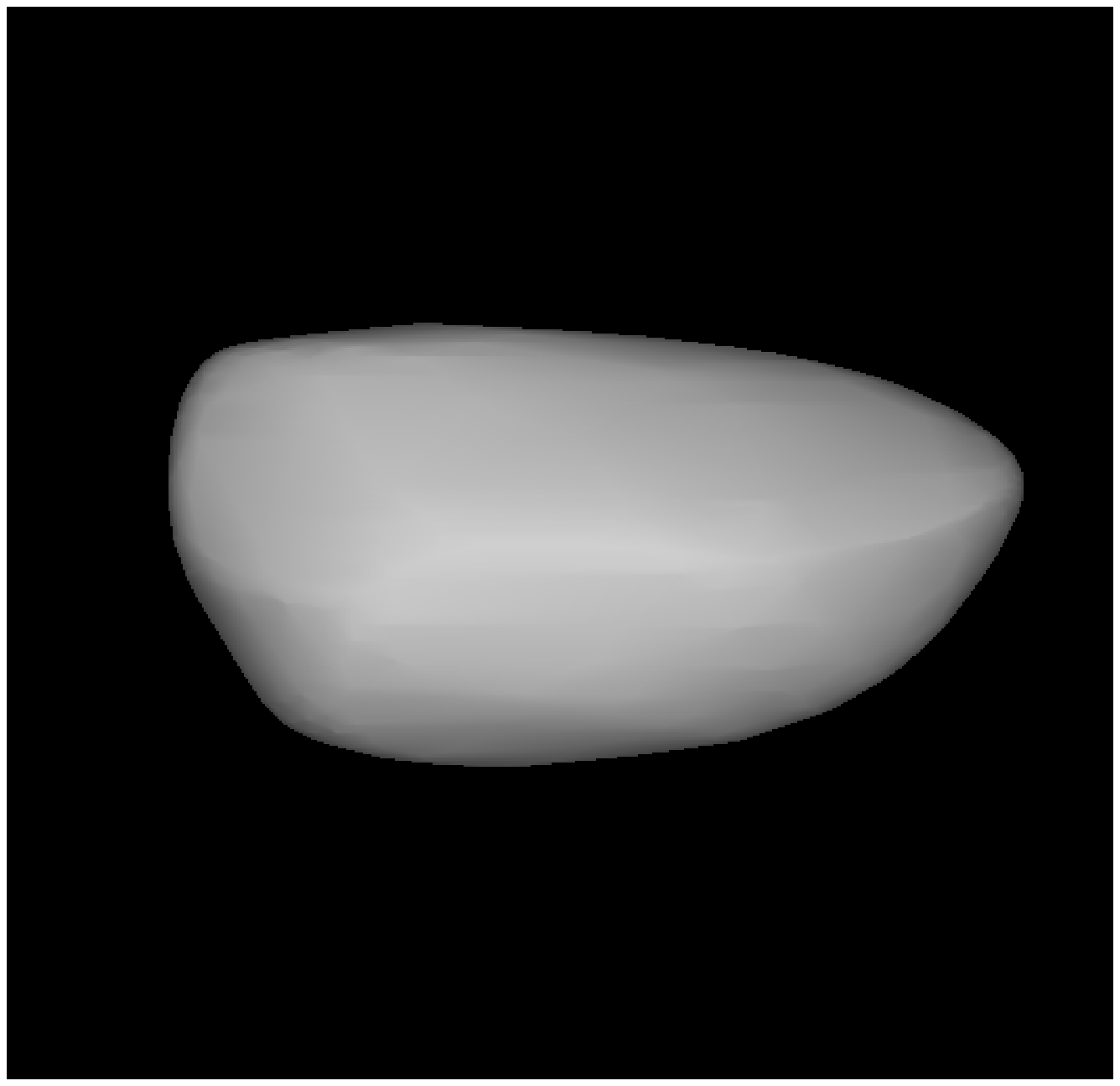}}
      \caption{Equatorial edge-on (top) and pole-on (bottom) images
        of the shape model.
       \label{fig:shape}}
    \end{center}
  \end{figure}

  The shape-model used here (Kaasalainen et al.\ \cite{kaasalainen05})
  is an update and refinement of the model presented in Kaasalainen
  et al.\ (\cite{kaasalainen03}). The model accommodates
  new photometric observations from December 2003 to September 2004,
  as well as some 2001 data additional to the 2000-2001 apparition
  dataset presented in Kaasalainen et al.\ (\cite{kaasalainen03}).
  The long time-line of the updated dataset allowed accurate
  period determination for \object{Itokawa}, and a refined pole and shape
  estimate.
  All parts of Itokawa's surface were well visible during the two
  apparitions; however, the long period precluded fully covered rotational
  phases for single lightcurves.
  Calibrated photometry allowed the determination of Itokawa's
  solar phase curve for a wide range of solar phase angles. The refined
  rotation parameters are $\beta=-89^\circ\pm 5^\circ$,
  $\lambda=330^\circ$ for the ecliptic latitude and longitude of
  the pole, and $P=12.13237\pm 0.00008$\,h for the sidereal period.
  Fig.~\ref{fig:shape} shows equatorial edge-on and pole-on images
  of the shape model. The model agrees well with the radar-based
  one (Ostro et al.\ \cite{ostro05}).

\section{Thermophysical modelling on the combined dataset}
\label{sec:mod}

  \begin{figure}[h!tb]
    \begin{center}
      \rotatebox{90}{\resizebox{!}{\hsize}{\includegraphics{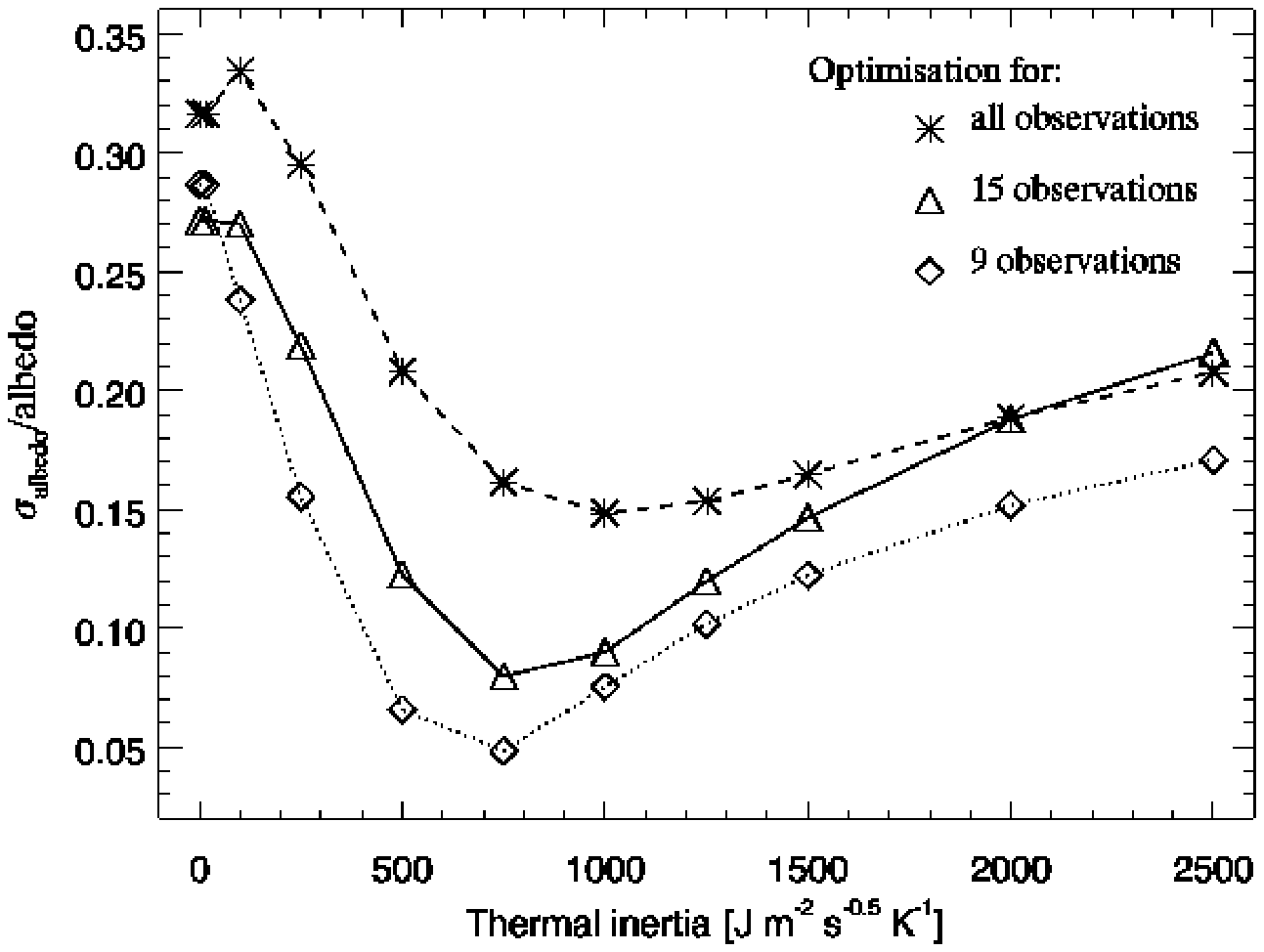}}}
      \caption{Thermal inertia optimisation process for the 
               individual TPM albedos and their standard deviation,
	       using all 20 individual observations (dashed line),
	       a subset with 15 observations (solid line)
	       and a subset of 9 observations (dotted line). 
       \label{fig:sig_albedo}}
    \end{center}
  \end{figure}

  \begin{figure}[h!tb]
    \begin{center}
      \rotatebox{90}{\resizebox{!}{\hsize}{\includegraphics{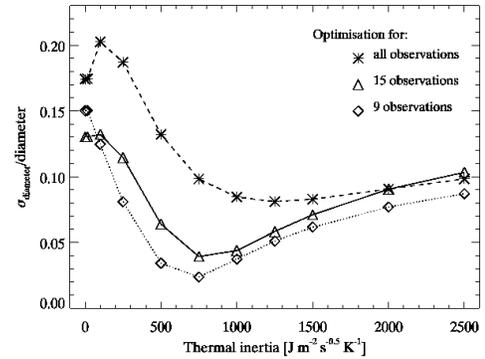}}}
      \caption{Thermal inertia optimisation process for the 
               individual TPM diameters and their standard deviation,
	       using all  20 individual observations (dashed line),
	       a subset with 15 observations (solid line)
	       and a subset of 9 observations (dotted line).
       \label{fig:sig_diameter}}
    \end{center}
  \end{figure}
  
  Our data set is very homogenous with respect to the used instrument
  settings, observing technique, data reduction and calibration scheme. 
  The data cover a wide range of
  different observing geometries, including before and after
  opposition, at different wavelengths and rotational phases.
  Therefore, it was possible to adjust the thermal inertia and
  beaming parameters to see how these variations influence the calculation
  of the radiometric diameter and albedo solutions.
  The investigations of the resulting diameters and albedos give clues about
  the optimal model parameters. Only specific model parameters will allow
  that e.g.\ the data taken at very different phase angles before and after
  opposition will result in the same radiometric diameter and albedo solutions.
  Another quality criteria is that the resulting diameter and albedo values
  show no trends with wavelengths or rotational phase.
  The goal was to find the best diameter/albedo solution with the smallest
  standard deviations which fits all 20 measurements.
  A similar procedure was already used by M\"uller \& Lagerros (\cite{mueller98};
  \cite{mueller02})
  for several main-belt asteroids. For large, regolith covered asteroids,
  the least-square process gave typical thermal inertias of
  10-15\,J\,m$^{-2}$\,s$^{-0.5}$\,K$^{-1}$ and beaming parameters of
  $\rho$=0.7 and $f$=0.6 (M\"uller et al. \cite{mueller99}).

  As a first step, we tried to find out how the thermal inertia 
  influences the determination of the radiometric
  diameter/albedo solutions. For observations taken at very large phase
  angles the thermal inertia is the most important parameter for
  a consistent diameter and albedo determination. This is due to the
  non-zero temperature of the large terminator which contributes
  significantly to the disk-integrated flux.  
  The thermal inertia $\Gamma$ was varied in a physically meaningful range between
  0 and 2500\,J\,m$^{-2}$\,s$^{-0.5}$\,K$^{-1}$. Where $\Gamma$=0 describes
  a surface in instantaneous equilibrium without any thermal conduction
  into the sub-surface, while $\Gamma$=2500\,J\,m$^{-2}$\,s$^{-0.5}$\,K$^{-1}$ corresponds to a highly
  conductive solid granite surface without any dust regolith. The Moon, with
  it thick highly insulating dust layer, has a thermal inertia of
  39\,J\,m$^{-2}$\,s$^{-0.5}$\,K$^{-1}$ (Keihm \cite{keihm84}).

  This wide range of thermal inertias has no big impact on the
  resulting weighted mean diameter and albedo values: A $\Gamma$=0
  would give weighted mean values of $D_{\rm{eff}}$=0.30\,km and $p_{\rm{V}}$=0.24,
  while a $\Gamma$=2500 would give $D_{\rm{eff}}$=0.33\,km and $p_{\rm{V}}$=0.19.
  But the standard deviation of the 20 diameter (or albedo) values
  changes enormously (more than a factor of 2)
  for different thermal inertias.
  Figures~\ref{fig:sig_albedo} and \ref{fig:sig_diameter}
  illustrate this effect on basis
  of the $\sigma_{\rm{albedo}}/albedo$ and $\sigma_{\rm{diameter}}/diameter$
  values. A thermal inertia of
  $\Gamma$=1000\,J\,m$^{-2}$\,s$^{-0.5}$\,K$^{-1}$ would therefore give
  the best match with our complete observational data set
  (dashed lines).
  We repeated the whole optimisation process with the highest 
  quality data only (excluding the data from April, 9th, 2001,
  M.\ Delbo, priv.\ comm.). The resulting best thermal inertia
  would then be at around 750\,J\,m$^{-2}$\,s$^{-0.5}$\,K$^{-1}$
  (solid line). A last robustness check with only 9
  observations (dotted line) confirmed this solution.

  In a second iteration, we tested the beaming model, parameterised
  by $\rho$, the r.m.s.\ of the
  surface slopes, and f, the fraction of surface covered by craters.
  Both parameters were kept variable between 0.1 and 0.9 (see also Dotto et al.\
  \cite{dotto00}).
  However, the effects with $\rho$ and $f$ are not as dramatic, mainly
  because most of our observations were taken at large phase
  angles where the beaming does not play an important role.
  We could not find a clear minimum in the $\sigma_{\rm{albedo}}/albedo$
  values in the $\rho$-$f$ plane. Some good solutions disappeared
  again when we checked for robustness by using various subsets
  of the observational data. As a conclusion from all optimisation
  runs we can only say that the very smallest values (0.1-0.3)
  are very unlikely for the beaming parameters $\rho$ and $f$
  for \object{Itokawa}, all other values still seem to be in
  agreement with our data set. We accepted therefore the default
  beaming values, $\rho=0.7$ and $f=0.6$, which were derived
  for main-belt asteroids (M\"uller et al.\ \cite{mueller99}).
  
  As result of this optimisation process we accepted the following 
  values:
  
  \begin{center}
  \begin{tabular}{lcll}
  $\Gamma$ & $=$ & 750   & J\,m$^{-2}$\,s$^{-0.5}$\,K$^{-1}$, thermal inertia \\
  $\rho$   & $=$ & 0.7   & r.m.s. of the surface slopes \\
  $f$      & $=$ & 0.6   & fraction of surface covered by craters \\
  \end{tabular}
  \end{center}

  Using the above parameters we derived the weighted mean values
  for $D_{\rm{eff}}$ and $p_{\rm{V}}$ together with the standard
  deviations\footnote{Accepting
  the results from the 15 highest quality data only
  decreases the standard deviations significantly.}:
  
  {\bf
  \begin{center}
  \begin{tabular}{lcl}
  $D_{\rm{eff}}$   & $=$ & 0.32$\pm$0.03\,km  ($\pm$0.01\,km) \\
  $p_{\rm{V}}$     & $=$ & 0.19$\pm$0.03 ($\pm$0.01)\\
  \end{tabular}
  \end{center}
  }
    
\section{Thermophysical modelling of the new dataset}
\label{sec:modnew}

\begin{figure}[h!tb]
    \begin{center}
      \rotatebox{90}{\resizebox{!}{\hsize}{\includegraphics{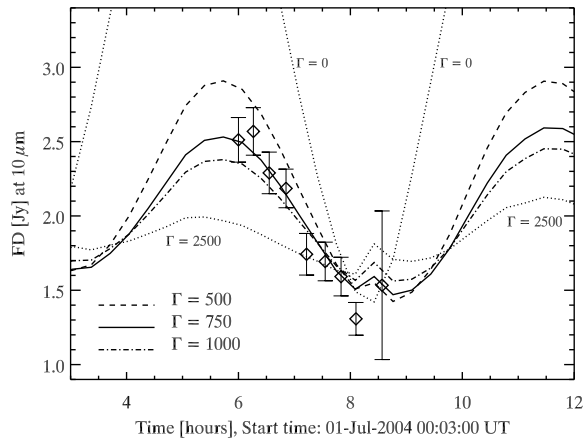}}}
      \caption{Predicted thermal lightcurve at 10.0\,$\mu$m for the
               time period around the July 1st, 2004 observations. The
	       original measurements were "transported" to the 10.0\,$\mu$m
	       wavelength via our TPM solution. Predictions and measurements
	       are shown with their absolute values, no shifting or scaling
	       in time or flux have been done. 
       \label{fig:tlc_10_750}}
    \end{center}
  \end{figure}

  We also used the TPM on basis of the established parameters to "transport"
  or normalise the observed fluxes to a reference wavelength of 10.0\,$\mu$m.
  Here we concentrated on the third dataset which has a 2.5\,hour
  coverage of the 12.1\,h rotation period. 
  For each of the measured values in Tbl.~\ref{tbl:obsres} we calculated
  the corresponding $D_{\rm{eff}}$ and $p_{\rm{V}}$ values (for
  $\Gamma$ = 750\,J\,m$^{-2}$\,s$^{-0.5}$\,K$^{-1}$, $\rho$=0.7, $f$=0.6).
  This was then used again 
  to predict the 10.0\,$\mu$m brightness for the given epoch
  (see diamond-symbols in Fig.~\ref{fig:tlc_10_750}).
  A weighted average $D_{\rm{eff}}$ and $p_{\rm{V}}$ (of dataset \#3) was
  then taken to predict the thermal lightcurve
  (solid, dashed, dashed-dotted, and dotted lines in
  Fig.~\ref{fig:tlc_10_750}), based on the Kaasalainen shape and
  spin-vector model and different thermal inertia values.

  Figure~\ref{fig:tlc_10_750} confirms that the thermal
  inertia has to be somewhere in the range between 500 and
  1000\,J\,m$^{-2}$\,s$^{-0.5}$\,K$^{-1}$. It also demonstrates that
  the implementation of the shape model in combination with
  the spin vector and zero points in time and phase yields
  consistent results. Our observational data cover the
  rotational phases between 354.9$^{\circ}$ (14-Mar-2001 05:50:00 UT)
  and 71.1$^{\circ}$ (01-Jul-2004 08:37:00 UT).

  
\section{Discussion}	\label{sec:discussion}

  All optimisation steps work well under the assumption that the
  shape and spin vector solutions are of good quality and that the
  albedo is the same all over the asteroid surface, e.g.\ for large
  main-belt asteroids which have almost spherical shapes with a
  very homogeneous albedo distribution due to a thick dust regolith.
  The shape model of \object{Itokawa} matches the visual lightcurves taken
  at very large range of phase angles and different observing
  and illumination geometries (Kaasalainen et al.\ \cite{kaasalainen05}).
  Kaasalainen et al.\ (\cite{kaasalainen03}) detected no significant
  albedo variegations. Thus, both prerequisites are fulfilled
  and it was for the first time possible to extract thermal properties
  for such an elongated object.
  So far, this was only possible for large main-belt asteroids 
  (e.g. Spencer et al.\ \cite{spencer89}; M\"uller \& Lagerros
  \cite{mueller98}) and a few NEAs (e.g.\ Harris \& Davies \cite{harris99})
  with almost spherical shapes where the radiometric diameter/albedo
  solutions were not affected too much by shape effects or albedo deviations
  at certain epochs.
  
  \begin{figure}[h!tb]
    \begin{center}
      \rotatebox{90}{\resizebox{!}{\hsize}{\includegraphics{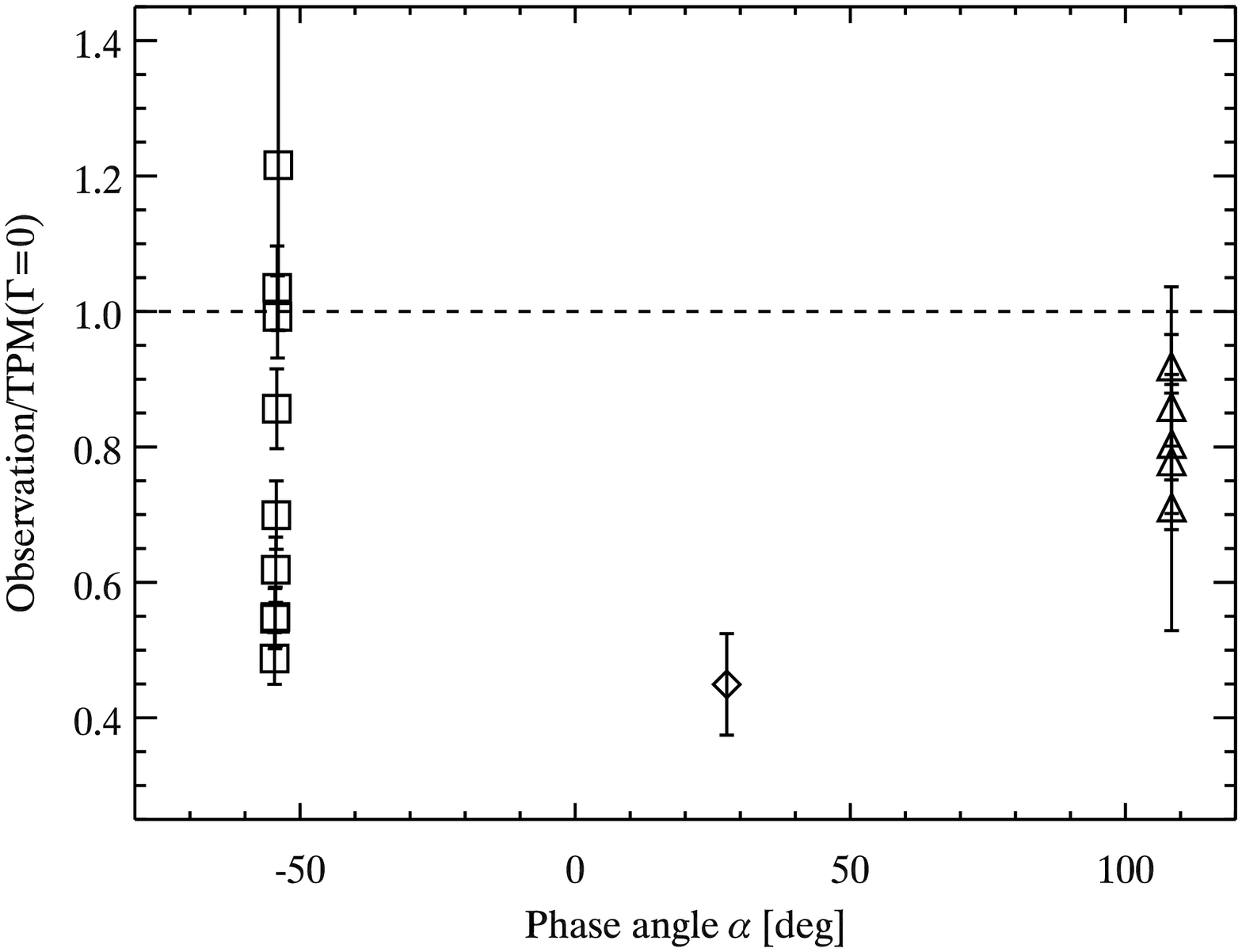}}}
      \rotatebox{90}{\resizebox{!}{\hsize}{\includegraphics{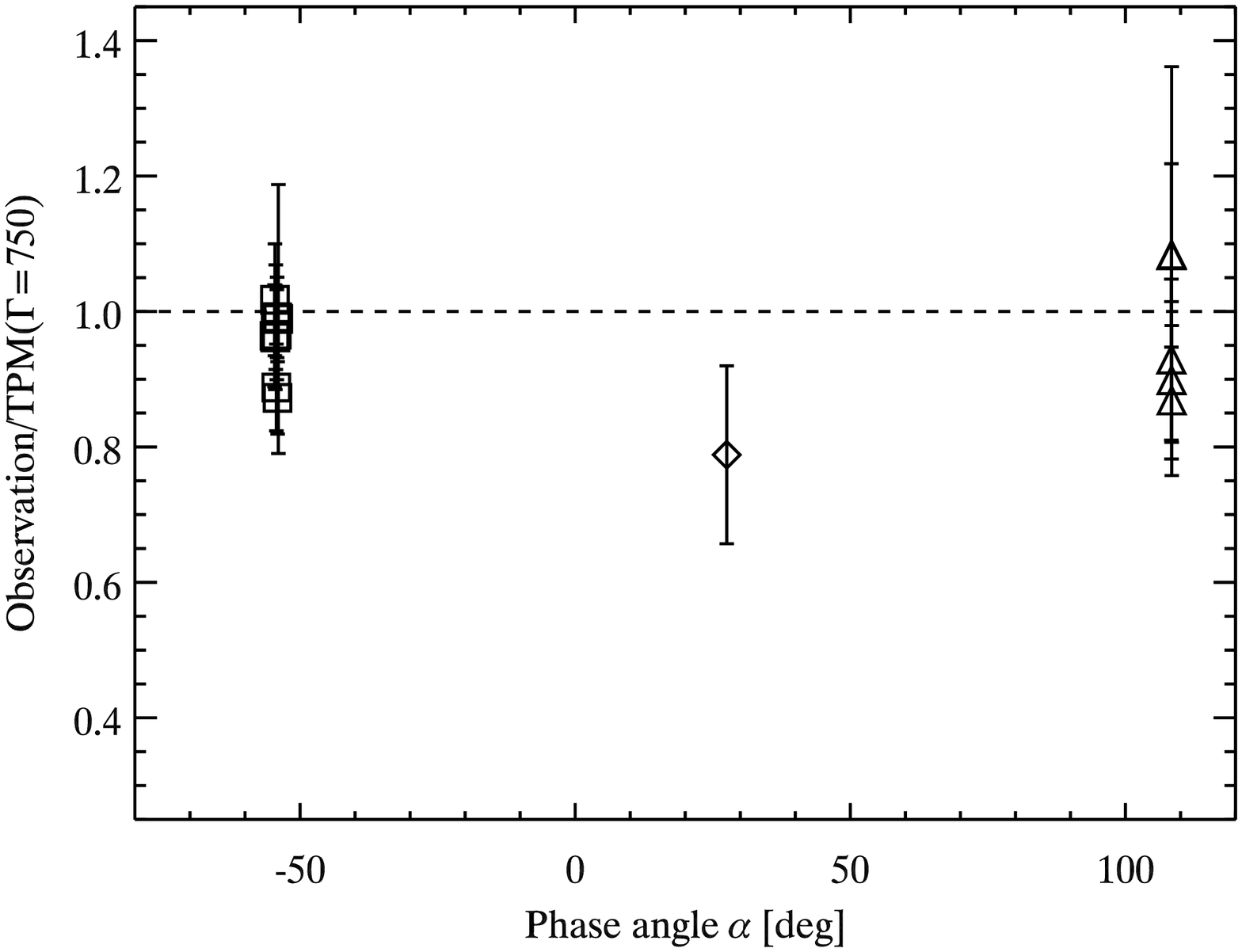}}}
      \caption{The observation/TPM ratios for a thermal inertia of 0 (top)
               and 750 (bottom). The high thermal inertia values eliminate
	       the trend with phase angle and reduce the scatter significantly.
       \label{fig:obsmod_alpha}}
    \end{center}
  \end{figure}

  \begin{figure}[h!tb]
    \begin{center}
      \rotatebox{90}{\resizebox{!}{\hsize}{\includegraphics{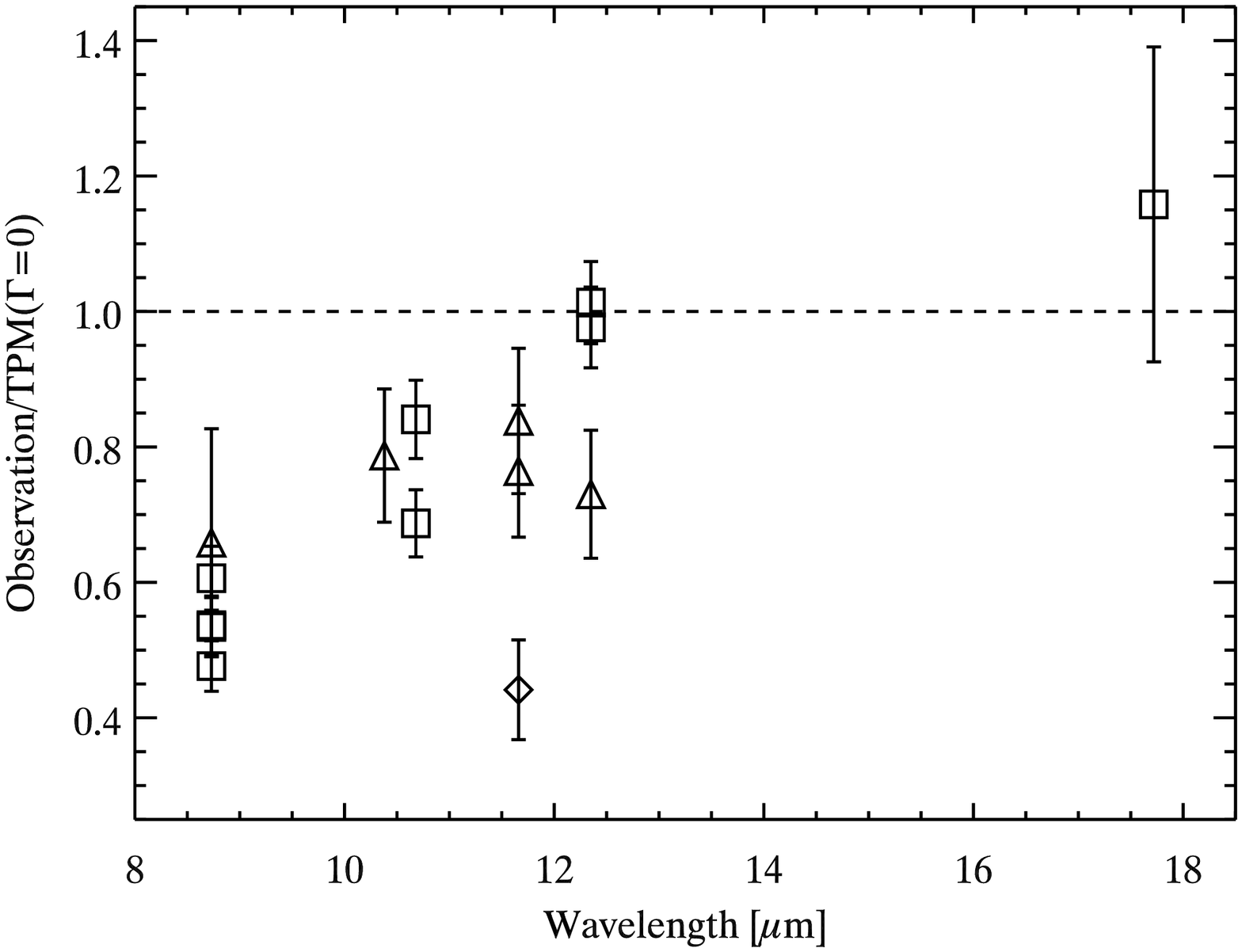}}}
      \rotatebox{90}{\resizebox{!}{\hsize}{\includegraphics{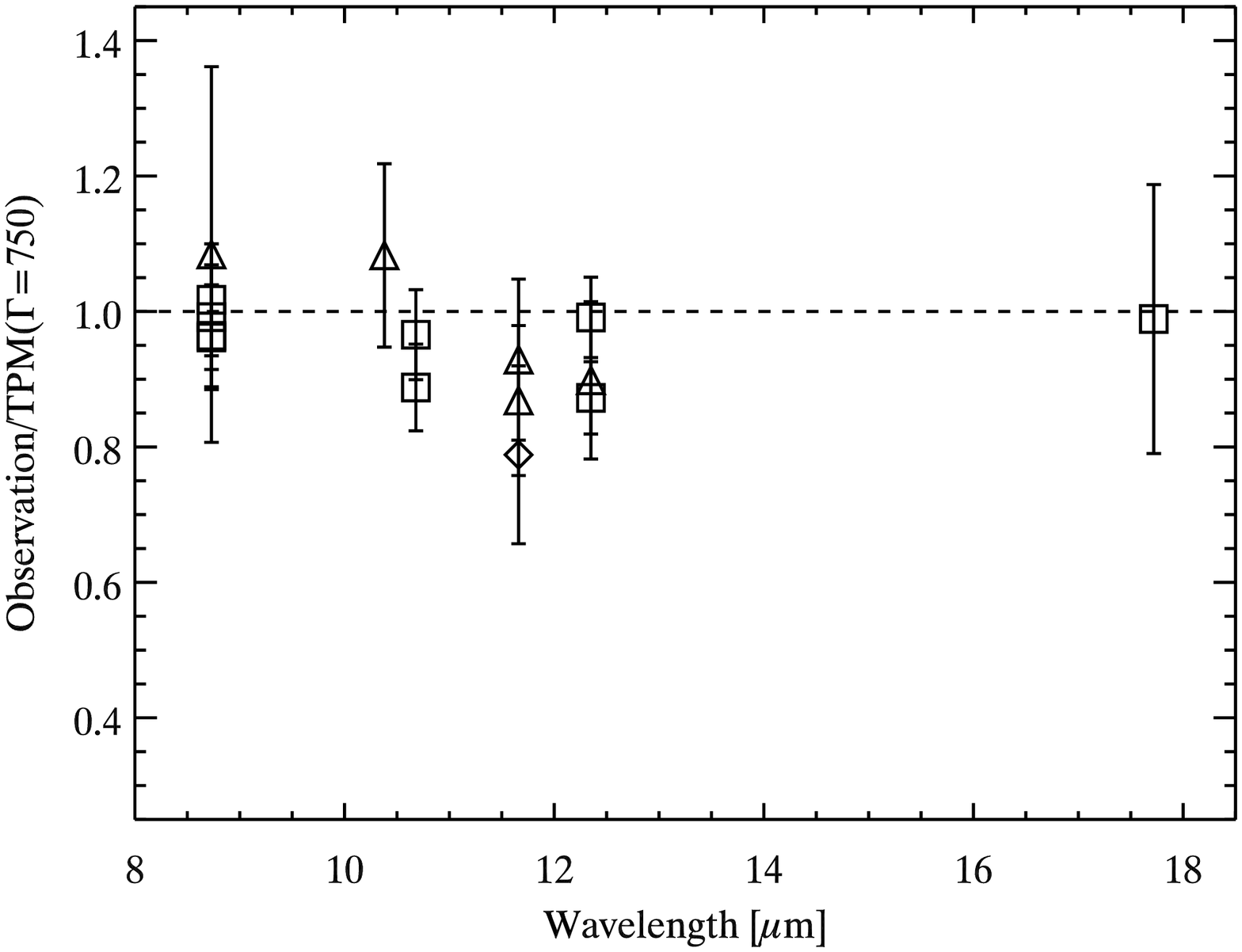}}}
      \caption{The observation/TPM ratios for a thermal inertia of 0 (top)
               and 750 (bottom). The high thermal inertia values eliminate
	       the trend with wavelength and reduce the scatter significantly.
       \label{fig:obsmod_lam}}
    \end{center}
  \end{figure}

  Another key element for a successful derivation of thermal parameters
  is the wavelength and phase angle coverage of the observational data set.
  In order to separate beaming and thermal inertia effects, it is necessary
  to have equal quality data at different phase angles.
  An often used indicator for reasonable model assumptions is the
  ratio plot "observation/model prediction" (e.g.\ M\"uller \& Blommaert
  \cite{mueller04a}).
  
  Figure~\ref{fig:obsmod_alpha} shows this "obs/mod"-ratio plot
  for the phase angles. Asymmetries in
  the phase angle plot (see Fig.~\ref{fig:obsmod_alpha}, top) confirm
  the retrograde sense of rotation (see e.g.\ M\"uller \cite{mueller02a} for
  further discussions) and show that the
  terminator has very different temperatures before ($\alpha > 0$)
  and after opposition ($\alpha < 0$),
  resulting in a poor modelling (large scatter) in the after opposition
  ratios.
  The TPM requires therefore a higher thermal inertia to match the
  observed fluxes (Fig.~\ref{fig:obsmod_alpha}, bottom) and to eliminate
  the phase angle asymmetry.
  
  Figure~\ref{fig:obsmod_lam} shows how the "obs/mod"-ratio
  varies with wavelengths. In case of a low thermal inertia (top),
  one can clearly see a trend of the ratio with wavelength.
  For the $\Gamma$=750-case, the scatter of the data points
  is much smaller and the trend with wavelength disappeared.
  Note, that the data points, which are considered 
  to be less reliable (Delbo, priv.\ comm.), are not shown for clarity
  reasons.
  
  Our derived albedo of $p_{\rm{V}}$ = 0.19$\pm$0.03 fits nicely within
  the established S-type albedo range of 0.21$\pm$0.07 (Ishiguro et al.\
  \cite{ishiguro03}). But this albedo value is strongly connected to
  the $H_{\rm{V}}$-value of 19.9\,mag. A change of 0.1\,mag in $H_{\rm{V}}$ to 19.8\,mag
  would increase the radiometric albedo by about 0.02, while the effective
  diameter would remain practically unchanged. Even very large modifications
  of the $H_{\rm{V}}$-value by e.g., 0.5\,mag to $H_{\rm{V}} = 19.4$\,mag
  ($H_{\rm{R}}=19.0$\,mag from Nishihara et al.\ \cite{nishihara05} and
  $(V-R) = 0.4$\,mag from Lowry et al.\ \cite{lowry05}) would 
  lead to a marginally increased $D_{\rm{eff}} = 0.33$\,km.
  But in this case, the directly connected
  albedo would increase to $p_{\rm{V}}=0.29$. Based on these considerations,
  we give a final solution of $p_{\rm{V}} = 0.19^{+0.11}_{-0.03}$ to account
  for the different published $H_{\rm{V}}$ values.
  This shows that the size determination is closely coupled to the quality
  of the thermal photometry, while the albedo depends much more on the
  properties from the reflected light analysis, i.e., the values
  in the H-G magnitude system. It should also be noted here that
  the H-G concept is a badly defined convention for irregular bodies.
  The $H_{\rm{V}}=19.9$\,mag is the best possible solution in the context
  of the Kaasalainen et al.\ (\cite{kaasalainen05}) shape model.
  Nishihara et al.\ (\cite{nishihara05})
  also determined a larger $G$-value of 0.25 (instead of the 0.21 used here). 
  But this difference would no be noticeable in the radiometric results.
  We also checked the robustness of our $\Gamma$-solution against uncertainties
  in the $H-G$ values. But taking the Nishihara et al.\ (\cite{nishihara05})
  $H-G$ values ($H = 19.4$\,mag, $G = 0.25$) leads to the same, very pronounced
  minimum at around 750\,J\,m$^{-2}$\,s$^{-0.5}$\,K$^{-1}$
  in the $\sigma_{\rm{albedo}}/albedo$ picture
  (see Fig.~\ref{fig:sig_albedo}).
    
  The derived radiometric value $D_{\rm{eff}}$ is the diameter
  of an equal volume sphere, based on the Kaasalainen-shape and spin-vector
  model. A rotating ellipsoid approximation can be described by
  the absolute sizes $2a$, $2b$ and $2c$, corresponding roughly to
  the x-, y- and z-dimensions of Fig.~\ref{fig:shape} (longest
  extension, hight in bottom and hight in top image):

  \centerline{
  $2a$ = 0.52$\pm$0.05\,km;
  $2b$ = 0.27$\pm$0.03\,km;
  $2c$ = 0.23$\pm$0.02\,km;
  }

  Note however, that for very general shapes, like the Itokawa-shape by Kaasalainen
  et al.\ (\cite{kaasalainen05}), the $a/b$ and $b/c$ ratios are not
  well defined.

  The radar observations (Ostro et al.\ \cite{ostro05}) resulted
  in a size estimate of 594$\times$320$\times$288\,m ($\pm$10\%),
  based on the same lightcurve-based spin vector by Kaasalainen
  et al.\ (\cite{kaasalainen05}). The axis ratios of the lightcurve and
  radar shape models agree and have approximate values of $a/b$=1.9
  and $b/c$=1.1.\footnote{Note here again that the dimensions are for a
  rough and not well-defined ellipsoid representation rather than, e.g.,
  for the largest extents.} But the effective diameter $D_{\rm{eff}} = 2(abc)^{1/3}$
  of the radar solution is about 15\% higher. We tried to use the
  radar effective
  diameter of about 0.38\,km to fit the observed highest quality
  data from July 1, 2004. Only with very unrealistic assumptions
  of $p_{\rm{V}}$=0.5 and a thermal inertia $\Gamma$=5000\,J\,m$^{-2}$\,s$^{-0.5}$\,K$^{-1}$
  the TPM predictions would provide an acceptable match with
  the observed fluxes. Taking the stated radar uncertainty
  of $\pm$10\% into account, we conclude that the radar sizes are
  overestimated by a few percent.
    
  Sekiguchi et al.\ (\cite{sekiguchi03}) derived through the
  NEATM (Harris \cite{harris98}) radiometric diameter
  and albedo values which agree within the specified error bars with
  our solution. The NEATM results for the NEA 2002~NY40 (data at
  one phase angle only) also 
  compare well with the TPM predictions (M\"uller
  et al. \cite{mueller04b}). But for data sets covering very 
  different phase angles, as it is the case here, the NEATM requires
  different beaming parameters (Delbo et al.\ \cite{delbo03}),
  while the TPM can explain all observed data points with one
  set of physical and thermal parameters. The TPM beaming model,
  parameterised by $\rho$ and $f$, can handle the very different
  illumination geometries without artificial correction factors.
  Additionally, the NEATM cannot explain the before/after opposition
  (or morning/evening) effect, unless the beaming parameters are adjusted
  differently before and after opposition. A detailed comparison
  with the NEATM was therefore not performed.
  
  We also tried to determine surface roughness
  properties, described in the TPM by $\rho$ and $f$.
  It turned out, that the TPM predictions for our given large
  phase angles are almost independent of these beaming parameters.
  This is in agreement with the fact that the effect only plays
  a role at small phase angles where mutual heating within the
  crater structures produce an enhanced amount of infrared
  flux (as compared to a smooth surface). Our phase angle
  coverage therefore does not allow to draw any conclusions
  on the surface roughness, crater structures or r.m.s.\
  values of the surface slopes. Additional data at small phase
  angles close to opposition are required for such investigations.
  
  We also compared the difference between using a spherical shape model 
  (together with the true spin vector solution under the given observing
  geometries in combination with the TPM) and using the
  Kaasalainen shape model. Assuming a spherical shape gives in fact very
  similar mean (or weighted mean) diameter and albedo values, but the
  scatter between the 20 derived albedo values is about 50\% larger.
  The derived diameter and albedo values are then dependent on the
  rotational phase and to a certain extent also on the aspect angle.
  Additionally, the effective diameter and albedo one obtains by simply
  averaging our observations is not significantly different than that derived
  from the rotationally resolved observations, since they span a range more
  or less uniformly from maximum to minimum of the lightcurve, as shown
  in Fig.~\ref{fig:tlc_10_750}.
  The observations in July 2004 allowed to determine a 10.0\,$\mu$m
  lightcurve which is perfectly matched by the Kaasalainen et al.\
  (\cite{kaasalainen05}) shape and spin vector model together
  with the TPM and a thermal inertia of
  750\,J\,m$^{-2}$\,s$^{-0.5}$\,K$^{-1}$. In general, measurements
  taken at relatively large phase angles after opposition
  (here $\alpha = -54^{\circ}$), where the terminator is still warm
  (for an object with retrograde rotation),
  can be considered as key observations to determine thermal
  properties of NEAs. The only important
  point is a sufficient coverage of the rotational phases.
  The measurements before opposition are much more
  influenced by the actual illumination and observing geometry and only
  in second order by the contribution from the cold terminator.
  
  The thermal inertia is defined as
  $\Gamma = \sqrt{\kappa_{\rm{s}} \rho_{\rm{s}} c_{\rm{s}}}$,
  with $\kappa_s$ being the thermal conductivity, $\rho_{\rm{s}}$ the density
  and $c_{\rm{s}}$ the heat capacity of the surface material.
  A dust layer on the \object{Itokawa} surface, like the one on the Moon,
  with typical Moon-like $\kappa_{\rm{s}}$ and $c_{\rm{s}}$ values (Keihm \cite{keihm84})
  would require a density several hundred times higher than the 1250\,kg m$^{-3}$
  for the Moon to account for the derived thermal inertia.
  On the other hand, combining the derived high thermal inertia with
  the S-type bulk density of \object{Ida} (Belton et al.\ \cite{belton95})
  of 2600\,kg m$^{-3}$ and the specific heat of Granite $c_{\rm{s}}$=800\,J kg$^{-1}$ K$^{-1}$
  the $\kappa_{\rm{s}}$ value would be in the order of 0.3\,W m$^{-1}$ K$^{-1}$.
  This seems to be a reasonable conductivity for a porous stony material.
  The porosity itself can be determined from the assumed bulk density
  of 2600\,kg m$^{-3}$ in combination with an anhydrous 
  ordinary chondrite surface composition (Ishiguro et al.\ \cite{ishiguro03}):
  $p = (1 - \rho_{\rm{bulk}}/\rho) \sim 0.3$ (Belton et al.\ \cite{belton95}).
  In conclusion, the mid-IR data are in very good agreement with the
  assumption of a bare rock surface. A thick dust regolith can be 
  excluded as well as a metallic surface which would have a $\Gamma$-value
  above 10\,000\,J\,m$^{-2}$\,s$^{-0.5}$\,K$^{-1}$ and consequently produce
  a very small thermal lightcurve amplitude (see Fig.~\ref{fig:tlc_10_750}).
  
  The total mass of \object{Itokawa} is directly connected to the
  assumed bulk density and the determined volume of the Kaasalainen-shape
  model via the $D_{\rm{eff}}$ value (full uncertainty range of both quantities
  has been applied):
  
  \centerline{
  $M = \rho_{\rm{bulk}} \cdot Volume = 2600\,\frac{kg}{m^3} \cdot \frac{4}{3} \pi (\frac{D_{\rm{eff}}}{2})^3 =
  4.5^{+2.0}_{-1.8} \cdot 10^{10}\,kg$
  }
  
  
\section{Conclusion}    \label{sec:conclusion}

 The example of \object{Itokawa} shows the potential of
 the TPM applications for NEAs. State-of-the-art shape models from radar
 and lightcurve inversion techniques can be used for sophisticated
 thermo-physical investigations. In fact, the Itokawa case was the
 first implementation of Kaasalainen-shape models in the context
 of the TPM by Lagerros
 (\cite{lagerros96}; \cite{lagerros97}; \cite{lagerros98}).
 The TPM allows the combination
 of observational
 data taken at different observing and illumination geometries
 and wavelengths. No artificial
 fitting parameters are necessary to explain the spectral
 energy distributions nor the thermal behaviour with phase angle.
 And the thermal lightcurve is a "normal" output product for
 objects with known sizes and shapes.
 
 The Hayabusa mission will characterise \object{Itokawa}'s properties
 with high reliability. Our derived properties can therefore be
 compared and the TPM be verified. This project will establish the
 "ground truth" for future NEA TPM applications.
 The experience with the thermal data are very important for
 the planning of future observing campaigns of NEAs. Depending
 on the availability of a target, the wavelengths, the phase
 angles and the rotational phases can be selected in such a way
 that the thermophysical characterisation benefits most. E.g.\
 if the surface roughness properties are of interest, one would
 have to include observations close to opposition. If thermal
 inertia is important, certain phase angles and/or a significant
 coverage of the thermal lightcurve are key ingredients for
 a successful study.

\begin{acknowledgements}
   We would like to thank F.\ Hormuth
   for his support in the data analysis of the
   TIMMI2 observations and J.\ Lagerros for his
   modifications in the TPM code to allow a proper
   use of the Kaasalainen shape models. M.\ Delbo
   supported our re-evaluation of the TIMMI2 data
   of his thesis work.
\end{acknowledgements}



\end{document}